# Optics Letters

# Low-power optical beam steering by microelectromechanical waveguide gratings

Carlos Errando-Herranz,[1,*] Nicolas Le Thomas,[2,3] and Kristinn B. Gylfason[1]

[1]Department of Micro and Nanosystems, KTH Royal Institute of Technology, Malvinas väg 10, SE-100 44 Stockholm, Sweden
[2]Photonics Research Group, INTEC Department, Ghent University-imec, Technologiepark-Zwijnaarde 15, 9052 Ghent, Belgium
[3]Center for Nano- and Biophotonics, Ghent University, Technologiepark-Zwijnaarde 15, 9052 Ghent, Belgium
*Corresponding author: carloseh@kth.se

Optical beam steering is key for optical communications, laser mapping (lidar), and medical imaging. For these applications, integrated photonics is an enabling technology that can provide miniaturized, lighter, lower-cost, and more power-efficient systems. However, common integrated photonic devices are too power demanding. Here, we experimentally demonstrate, for the first time, to the best of our knowledge, beam steering by microelectromechanical (MEMS) actuation of a suspended silicon photonic waveguide grating. Our device shows up to 5.6° beam steering with 20 V actuation and power consumption below the µW level, i.e., more than five orders of magnitude lower power consumption than previous thermo-optic tuning methods. The novel combination of MEMS with integrated photonics presented in this work lays ground for the next generation of power-efficient optical beam steering systems.   © 2019 Optical Society of America

https://doi.org/10.1364/OL.44.000855

Optical beam steering is required in a wide range of applications, such as high-speed optical communications [1], lidar for artificial vision [2], and medical imaging [3]. Traditionally, optical beam steering systems use electrical motors to tilt mirrors and scan a laser beam over a certain area, which suffer from large size and weight, cost thousands of USD, and consume watts of power [4]. These traditional systems are impractical for battery-driven robots, mobile phones, or drones, for *in vivo* optical coherence tomography (OCT) probes [3], and for miniaturized and low-cost space division multiplexing (SDM) [1]. More recently, optical beam steering systems have been scaled down by using microelectromechanical (MEMS) mirrors and gratings, resulting in significant reduction in cost and weight [5,6]. However, the parts of such systems (i.e., laser, scanning device, detector, and electronics) are still fabricated independently, and require costly assembly. Further miniaturization has the potential to provide smaller, lighter, and less power-consuming beam steering at a low cost—features that are required for the continued success of optical beam steering technologies [2].

Integrated photonics, and silicon photonics in particular, can potentially address these challenges by densely integrating photonic devices for beam steering and optical signal processing, optical sources, and detectors [7], with electrical processing and control [8]. This results in integrated photonics systems outperforming free-space optics not only in size and weight, but also in cost, integration density, and robustness.

Integrated photonics approaches to beam steering have focused mostly on optical phased arrays. An optical phased array consists of an array of emitters (usually grating couplers), resulting in a diffraction pattern in the far field highly dependent on the relative phases of the emitted waves. By tuning the relative phases of such waves using waveguide phase shifters, the output beam angle is tuned. These systems allow a very tight control over the beam shape and direction, and previous work has shown 1D steering [9], very high angular beam resolution 2D steering [10], and lidar measurements [11]. However, the commonly used thermo-optic phase shifters have an important drawback: very high power consumption. This is caused by the need for one phase shifter per emitter, requiring hundreds of devices packed in a tight chip space. This has led to power consumption on the order of watts (0.5 W for 1D steering in [9], and about 4 W in [10]), which necessitates active cooling and thus limits severely the applications of this technology. Recently, a low-power lidar based on reverse-biased electro-optic phase shifters achieved 2 µW power consumption per phase shifter, amounting to about 1 mW for the required array of 512 [12]. The optical loss was up to 4 dB per phase shifter.

A different approach to beam steering used a single thermo-optically tuned grating coupler, and achieved a limited steering of 2.7° while consuming 130 mW of electrical power in static operation [13]. Along this line, a recent approach combined thermo-mechanical actuation with thermo-optic tuning in a grating coupler, and used it to improve the efficiency of thermo-optic spectral tuning of the grating transmission [14].

MEMS tuning of photonic waveguide devices can provide more than five orders of magnitude lower power consumption compared to traditional tuning methods (sub-µW for MEMS [15] compared to 30 mW per thermo-optic phase shifter [16]), with prospects for upscaling photonic integrated circuit (PIC) technologies [17]. In our previous work, we demonstrated a MEMS tunable grating for on-chip optimization of light coupling between an optical fiber and a silicon waveguide by using vertical displacement of a grating embedded in a cantilever [18].

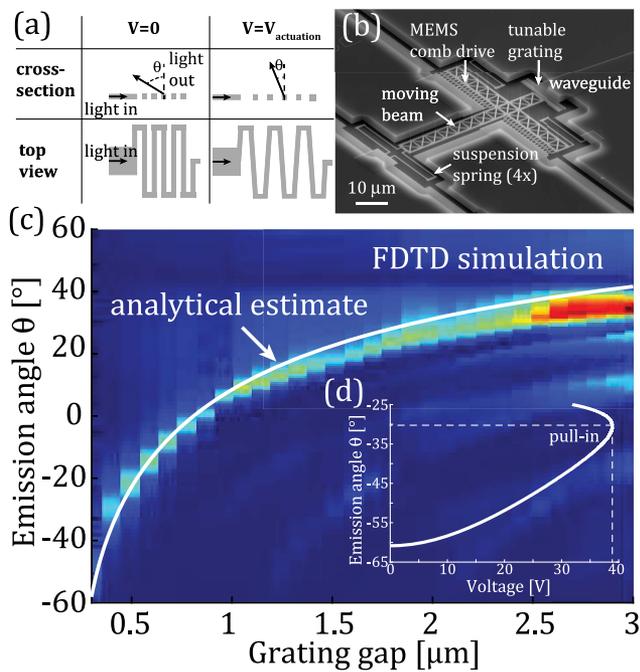

**Fig. 1.** (a) Schematic showing the working principle of our MEMS tunable grating before, and under actuation. (b) Scanning electron microscope (SEM) image of our device. The grating is part of a soft spring, stretched via a comb drive actuator, which changes the spacing between grating teeth. Note that the warped grating is due to early failure after actuation. (c) Simulation results (color: emitted light intensity) with overlaid analytical estimate (white line) of the effect of increased grating teeth separation on beam steering. (d) Analytical actuation estimate for a comb-drive actuated device.

In this Letter, we experimentally demonstrate, for the first time, low-power beam steering using a MEMS tunable waveguide grating. Our results show beam steering up to 5.6° at a wavelength of 1550 nm with actuation voltages below 20 V and sub-μW static power consumption.

Our beam steering device is based on changing the spacing between the teeth of a waveguide grating coupler using MEMS actuation. Figure 1(a) shows a schematic view of the device. We designed a suspended grating forming a folded spring, connected on one side to a waveguide taper for light coupling, and on the other side to a MEMS comb drive actuator. Horizontal actuation of the comb drive pulls and stretches the suspended grating, which changes the spacing between grating teeth, resulting in a change in the out-of-plane angular emission of the grating. Figure 1(b) shows a scanning electron microscope (SEM) image of a fabricated MEMS tunable grating.

An analytical estimate of the effect of varying the gap in a suspended grating can be obtained by using the standard grating equation assuming in-plane excitation, and estimating the effective grating index $n_{eff}$ as a weighted average between the effective refractive indices of the air gaps (width $g$) and silicon grating teeth (simulated $n_{wg} = 2.4$, width $d$):

$$\sin\theta = \frac{n_{eff} - \frac{\lambda}{d+g}}{n_{air}}, \quad \text{with} \quad n_{eff} = \frac{n_{wg}d + n_{air}g}{d+g}. \quad (1)$$

Thus, an equation relating the out-coupled light angle $\theta$ to the gap width $g$ was obtained. The white curve in Fig. 1(c) is the graph of that analytical relation, assuming a silicon grating tooth width $d = 300$ nm at $\lambda = 1550$ nm wavelength.

However, the analytical solution is only a rough approximation, due to the sub-wavelength scale of the structures involved. To get a more accurate prediction of the effect of varying grating spacing on the emission angle, optical simulations are required. We performed such simulations using a finite-difference time-domain solver (varFDTD, Lumerical Solutions), commonly used for grating coupler simulations in silicon photonics. We simulated the cross section of an air-cladded, suspended grating (tooth width $d = 300$ nm, device layer thickness $t = 220$ nm), including the under-etched buried oxide layer (thickness 2 μm) and the silicon substrate. The input waveguide was excited with the fundamental TE waveguide mode, and we investigated the out-coupled optical intensity in the far field for a grating with 15 teeth [Fig. 1(c)].

The mechanical design is based on a suspended comb drive actuator, with the attached tunable grating as an additional soft spring. The grating was designed using a tooth width of 300 nm, with 300 nm wide initial gaps, and grating width 20 μm abutting a waveguide taper with a wider end of 12 μm width [see Fig. 1(b)]. In our device, the change in gap between teeth is the total MEMS displacement divided by the number of teeth, and thus the number of teeth was chosen to be five to ease observation of beam steering even at short MEMS displacements. The grating design was 20 μm wide to minimize the in-plane angular variation from tooth to tooth under actuation (below 6° for 1 μm gap increase), so that along the central area of the grating, i.e., where the optical mode is concentrated, the effect of the variation in gap from tooth to tooth is negligible. The grating spring is designed to be soft (spring constant $k_{grat} = 3 \times 10^{-4}$ N/m), so that it is negligible in the MEMS actuation. The comb drive actuator is designed for fabrication on a standard silicon-on-insulator (SOI) 220 nm thick device layer, following [19]. The actuator uses four symmetric spring suspensions ($k_{springs} = 0.44$ N/m), designed as folded beams with beam width 300 nm, beam length 16 μm, and separation between beams of 3 μm. The force balance equation is

$$k_{springs}x = \epsilon_0 t N \left( \frac{1}{s} + \frac{w}{(D-x)^2} \right) V^2, \quad (2)$$

with $\epsilon_0$ the permittivity of vacuum, $t$ the device layer thickness, $N$ the number of comb finger pairs, $s$ the finger spacing, $w$ the width of each finger, $D$ the initial distance between the end of a finger and the beginning of the opposite one, $x$ the comb displacement, and $V$ the applied voltage. Our designed comb parameters are $t = 220$ nm, $N = 36$, $s = 400$ nm, $w = 300$ nm, and $D = 1.8$ μm.

The maximum displacement for our designed actuator is close to 900 nm, at a voltage of 40 V. Combining Eqs. (1) and (2) leads to the actuation curve for the MEMS tunable grating coupler in Fig. 1(d), with potential beam steering up to 30° at 40 V actuation.

The device was fabricated using the simple process presented in [20]. The process starts from a standard silicon photonic SOI substrate (220 nm Si device layer on 2 μm buried

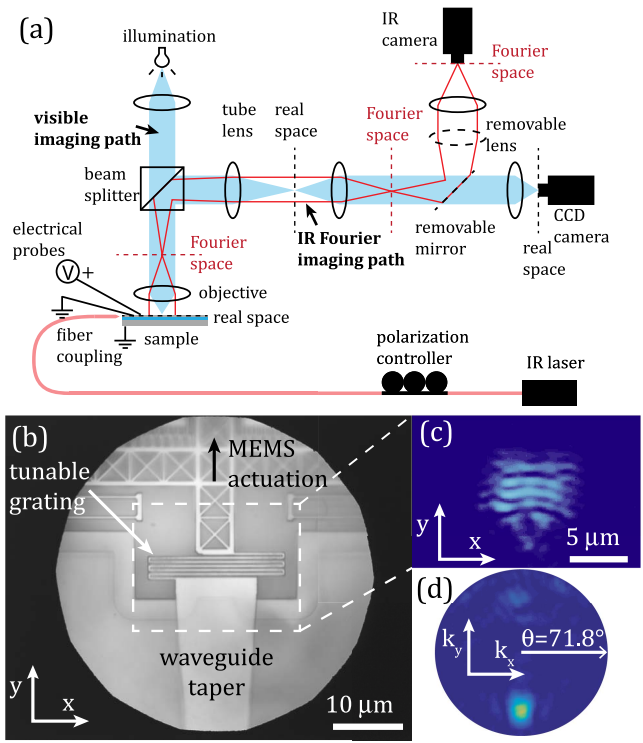

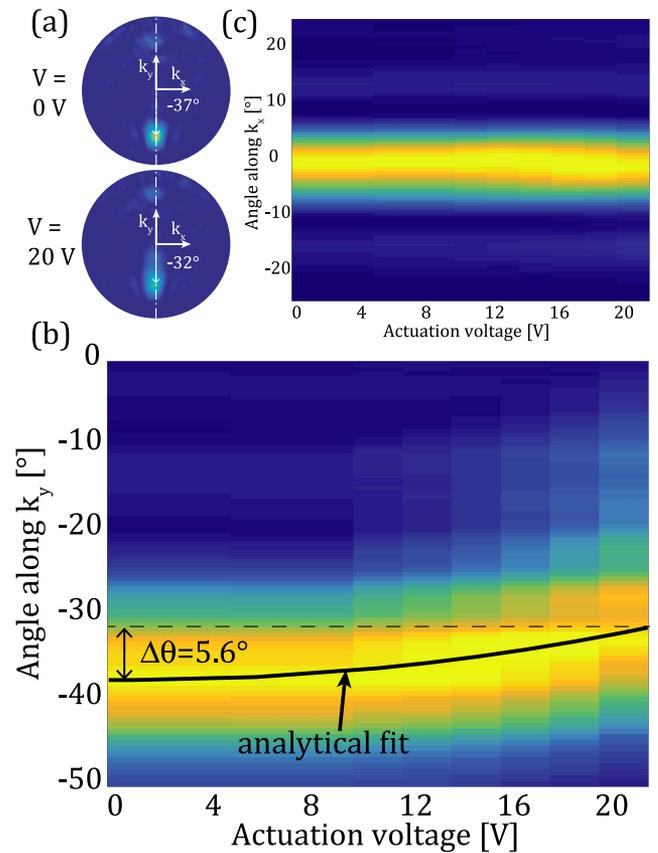

**Fig. 2.** (a) Experimental setup for measuring the angular emission from our devices by Fourier imaging. The mirror and Fourier lens at the rear end can be removed to view (b) the visible image of the grating (blue optical path). (c) By inserting the mirror, we image the IR emission of the grating onto the IR camera. (d) By inserting also the Fourier lens, we make the back focal plane visible, resulting in a k-space image of the grating emission on the IR camera (outlined red optical path).

**Fig. 3.** (a) k-space images showing the change in beam directionality with actuation. The white dashed lines show the cross-sectional axis used for plotting (b). (b) Beam steering for a range of actuation voltages along the MEMS actuation axis $k_x$. The results show up to 5.6° steering, using an analytical curve to fit the emission maxima (black curve). (c) Cross section along $k_x$ direction at the beam maxima for each voltage, showing negligible perpendicular beam distortion with actuation.

$SiO_2$), followed by two consecutive e-beam-lithography-defined silicon etching steps to pattern two silicon thickness levels. Then, a $SiO_2$ free-etch using hydrofluoric acid (HF, 50% aqueous solution) and critical point drying (CPD) results in the suspended MEMS structures [see Figs. 1(b) and 2(b)].

To characterize the beam steering, a Fourier imaging setup was used [21], and a schematic can be found in Fig. 2(a). This setup is based on an optical Fourier microscope using an objective with large numerical aperture (NA = 0.95), allowing measurement of beam angles up to 71.8° with respect to the normal.

Laser light at 1550 nm wavelength and 1 mW power is edge-coupled via a polarization controller through a standard optical fiber into the quasi-TE mode of the on-chip waveguides. On the chip, the light is guided to the tunable grating, which emits light out of plane into a large NA microscope objective, part of the Fourier imaging setup. In order to visually inspect the sample, visible light is sent into the objective using a beam splitter, and the reflected light is routed through a set of lenses for magnification, and finally onto a CCD camera [a visible light image is shown in Fig. 2(b)]. The infrared (IR) light emitted from our gratings follows the same optical path. However, optionally placing a mirror [labeled removable mirror in Fig. 2(a)] in front of the CCD camera reflects the light into an IR camera, and a real space image of the grating IR emission is formed, and can be seen in Fig. 2(c).

To obtain an image of the IR k-space, an additional lens can be placed in front of the IR camera [labeled removable lens in Fig. 2(a)], resulting in back focal plane (Fourier) imaging. An example of a k-space image of measured grating emission is shown in Fig. 2(d), which consists of a circle with its radius defined by the largest angular emission that the system can detect, i.e., 71.8°. The emission angle along the radial direction is defined by $r = \sin\theta$, with $r$ being the radial distance.

The substrate of the sample is electrically grounded via a copper plate, and two soft electrical probes, in direct contact with the silicon device layer, connect the comb drive electrodes to a voltage source for MEMS actuation. Actuation of the comb drive then results in pulling forces on the grating, changing the tooth spacing and resulting in beam steering.

Figure 3 shows our beam steering measurement results. The k-space images in Fig. 3(a) show the effect of increasing actuation voltage on the beam directionality along the MEMS actuation axis $k_y$ for 0 V and 20 V actuation. The absence of features other than beam steering, which would appear as different shapes in the k-space images, illustrates the absence of stray light or added noise in the k-space under MEMS actuation. A cross section along $k_y$ in Fig. 3(b) shows the evolution of the beam angle for a range of actuation voltages from 0 V to 20 V. Our device achieves $5.6 \pm 0.3°$ of beam steering with a full width at half maximum (FWHM) divergence of 14° along $k_y$ and 9° along $k_x$. The beam steering angles were obtained by fitting the maxima at each voltage to the actuation curve shape in Fig. 1(d), and the fit is plotted in Fig. 3(b), with the error value on beam steering angle calculated as one standard deviation. We can extract a tuning rate of $\Delta\theta/\Delta V = 0.56°/V$ from 10 V to 20 V. The cross section along the $k_x$ axis at the light intensity maxima, shown in Fig. 3(b), shows minor distortions in the direction perpendicular to the actuation, which can be originated from non-parallel displacement of the grating teeth. The efficiency of the device was estimated by simulating the suspended to unsuspended waveguide transition and the grating, and yielded <0.01 dB and 35% efficiency (varying between 30% and 40% with actuation), respectively. The static power consumption was in the sub-μW range, below our measurement capabilities. The nanogram mass of our MEMS structure makes gravitational forces negligible, and sets the mechanical resonance frequency around 200 kHz, which is far from mechanical noise sources, and sets a limit to the steering speed. The maximum power consumption for the device, estimated as 200 kHz full charge–discharge cycles, is below 10 nW.

The large beam divergence is due to the high-index contrast, and thus the limited number of grating teeth contributing to the diffraction pattern. This can be overcome by designing a lower-refractive-index-contrast grating by, e.g., thinning down the silicon device layer, or by choosing a different low-index material platform such as silicon oxide or silicon nitride.

The limited actuation range, less than half of the theoretical prediction, is due to premature collapse of the MEMS actuator, stemming from the grating spring, and can be observed in the tunable grating area in Fig. 1(b). We believe there are two reasons for this effect: (i) large displacements of the grating result in local stiffening and softening, resulting in bending momenta and contact between grating teeth, and (ii) the asymmetry of the grating spring generates an off-axis horizontal force under actuation, resulting in lateral drift of the comb drive and contact between grating teeth. We believe these problems are not fundamental, and can be solved by (i) design of grating teeth joints to avoid strain concentration, and (ii) design of symmetric gratings, either by mirroring the current design, or by designing a suspended grating connected on the sides.

Compared to free-space optics, the presented technology is orders of magnitude smaller and lighter, lower in cost, less prone to mechanical noise, and requires very limited assembly. Integrated thermo-optic phased array systems have at least five orders of magnitude (limited by measurement, seven orders of magnitude based on our estimate) higher power consumption than our device, and suffer from thermal cross-talk problems, which our technology inherently avoids. Compared to elecro-optic tuning, our device features at least one order (more likely three orders) of magnitude lower power consumption. Furthermore, we achieve more than two times larger beam steering than previously reported thermo-optic tunable single gratings [13], with potential for large-angle tuning [see Fig. 1(d)] with future improvements in MEMS actuator design.

We have introduced, for the first time, MEMS tunable waveguide gratings for beam steering, and experimentally demonstrated beam steering up to 5.6°, with an actuation voltage below 20 V. Our results show more than twice the beam steering of previously reported thermo-optic tuning of waveguide gratings [13], and more than five orders of magnitude lower power consumption.

The optical beam steering technology presented here can provide the long-sought reduction in cost and power consumption necessary to extend artificial vision to battery-powered devices, including mobile phones or drones, to enable active probes for *in vivo* medical imaging, and to grow the optical communication bandwidth by SDM.

**Funding.** Vetenskapsrådet (VR) (621-2012-5364); H2020 European Institute of Innovation and Technology (EIT) (780283).

**Acknowledgment.** N. L. T. acknowledges the support of the MRP initiative at Ghent University through NB-Photonics.